\documentclass[aps,prd,twocolumn,nofootinbib,superscriptaddress,floatfix,longbibliography,colorlinks=true,linkcolor=blue,citecolor=blue,urlcolor=blue]{revtex4-2}

\usepackage{amsmath,amssymb,amsfonts,mathtools}
\usepackage{bm}
\usepackage{graphicx}
\usepackage{orcidlink}
\usepackage{hyperref}

\newcommand{\ar}{\mathrm{ar}}
\newcommand{\BH}{\mathrm{BH}}
\newcommand{\lz}{\ell_\zeta}

\begin{document}

\title{From arithmetic spectra to a quantum-corrected black hole geometry}

\author{Kimet Jusufi\,\orcidlink{0000-0003-0527-4177}}
\email{kimet.jusufi@unite.edu.mk}
\affiliation{Physics Department, University of Tetova, Ilinden Street nn,
1200 Tetova, North Macedonia}

\author{Ankit Anand\,\orcidlink{0000-0002-8832-3212}}
\email{anand@iitk.ac.in}
\affiliation{Department of Physics, Indian Institute of Technology Kanpur,
Kanpur 208016, India}

\date{\today}

\begin{abstract}
The Euler product of the Riemann zeta function is the partition function of a
free bosonic gas whose mode energies are the logarithms of the primes. We show
that this arithmetic gas, combined with the assumption that the entropy--geometry correspondence holds,
leads to a
quantum-corrected black-hole metric. The prime gas is a Hagedorn system. Its
entropy is linear in the energy, which is exactly what an entropy linear in the
horizon area requires, and the simple pole of the zeta function at $\beta=1$
fixes the coefficient of the logarithmic correction to the area law. The
nontrivial zeros cannot play this role: their level density grows only
logarithmically, and far too slowly to be extensive. Demanding that the
reconstructed geometry reduce to Schwarzschild at large radius then fixes the
map from arithmetic energy to horizon area and yields the closed-form metric
$f(r)=1-2GMr/(r^{2}+\lz^{2})$ with $\lz^{2}=\alpha G/\pi$. This describes a
two-horizon black hole with Reissner--Nordstr\"om horizon structure but no
Coulombic hair, a positive-energy anisotropic source obeying the null energy
condition, a softened central singularity, a bounded Hawking temperature, and a
cold extremal remnant that ends the evaporation. The same length scale follows
independently from requiring that the first law hold exactly with the corrected
entropy. The nontrivial zeros survive only as exponentially suppressed
log-periodic ripples in the area, which suggests a physical interpretation of the Riemann hypothesis
as the statement that arithmetic corrections to black-hole thermodynamics are
as small as they can be.
\end{abstract}

\maketitle

\section{Introduction}
\label{sec:intro}

Taking a logarithm turns arithmetic into statistical mechanics. Unique
factorization, $n=\prod_p p^{a_p}$, becomes additivity of energies,
$\ln n=\sum_p a_p\ln p$, so the multiplicative structure of the integers is an
additive spectrum built on the primes. The Riemann zeta function displays this
directly,
\begin{equation}
\zeta(s)=\sum_{n\ge1}e^{-s\ln n}=\prod_p\left(1-e^{-s\ln p}\right)^{-1},
\quad \Re s>1 ,
\label{eq:zeta}
\end{equation}
which identifies the primes as bosonic modes of energy $\epsilon_p=\ln p$ and
the integers as their multiparticle states. This ``primon gas'' or ``Riemann
gas'' has been studied as a thermodynamic system since Julia~\cite{Julia}, with
bosonic, fermionic and supersymmetric variants developed by
Spector~\cite{Spector} and Bakas and Bowick~\cite{BakasBowick}, and an
operator-algebraic formulation given by Bost and Connes~\cite{BostConnes}. Its
defining feature is the pole of $\zeta$ at $s=1$: the density of prime modes
grows exponentially, the canonical partition function diverges at $\beta=1$, and
the system has a limiting (Hagedorn~\cite{Hagedorn,AtickWitten}) temperature.

A second and logically independent spectrum is supplied by the nontrivial zeros
$\rho_n=\tfrac12+it_n$. The Hilbert--P\'olya conjecture holds that the $t_n$ are
the eigenvalues of a self-adjoint operator, a picture supported by the pair
correlation of Montgomery~\cite{Montgomery} and the numerics of
Odlyzko~\cite{Odlyzko}, and pursued dynamically by Berry and
Keating~\cite{BerryKeating}, spectrally by Connes~\cite{Connes}, and in concrete
quantum models by Sierra and Townsend~\cite{SierraTownsend}. The two spectra are
tied together by the explicit formula, which reads like a semiclassical trace
formula: the zeros play the role of energy levels and the primes that of
primitive periodic orbits of length $L_{p^m}=m\ln p$.

Black-hole thermodynamics supplies the other half of the problem. The
Bekenstein--Hawking entropy $S_\BH=A/4G$~\cite{Bekenstein,Hawking} requires a
Hilbert space of dimension $e^{A/4G}$, and the microstates behind it have been
identified only in special settings---string theory~\cite{StromingerVafa}, loop
quantum gravity~\cite{ABCK}---each of which also predicts a subleading
logarithmic term $S=A/4G+\alpha\ln(A/4G)+\dots$
\cite{KaulMajumdar,Carlip,Das,Sen,Solodukhin}. The coefficient $\alpha$ is a
sharp diagnostic: it depends on the scheme and the ensemble, and any candidate
microscopic description has to produce a definite value for it. It is therefore
natural to ask whether either arithmetic spectrum can carry gravitational
entropy, and if so which one. The question is usually left at the level of
suggestive numerology. Our aim here is to push it until it produces a metric.

Two ingredients make this possible. The first is that the required exponential
degeneracy is not generic. It is a property of the Euler product and not of the
zeros, and we show below that this distinction is decisive and survives second
quantization. The second is the entropy--geometry
correspondence~\cite{Anand,AnandTopo,JusufiAnand}, which reverses the usual
logic of general relativity: instead of prescribing a source and solving for the
metric, one prescribes a horizon entropy function $S(r)$ and reconstructs the
static spherically symmetric geometry that implies it, together with the
effective anisotropic fluid that sources it. Given an arithmetic entropy, this
turns a counting statement into a spacetime.\\

The methodological chain of the paper proceeds as follows. Starting from the probability weight $\ln p$, the arithmetic spectrum $\Omega_{\rm ar}$ is obtained via state counting. This spectrum then yields the energy-dependent entropy $S_{\rm ar}(E)$, the energy--area map $E_c(A)$, and the area entropy $S(A)$, which are ultimately used to reconstruct the metric function $f(r)$. Crucially, the arithmetic spectrum, the state counting, and the entropy are derived from the underlying statistical framework; by contrast, the energy--area map and the entropy--geometry relation are not derived but explicitly imposed as physical assumptions. The results are as follows. The
prime gas has $S_\ar\simeq E/T_H$. Requiring that the reconstructed geometry
reduce to Schwarzschild in the infrared fixes the arithmetic cutoff to
$E_c=T_HA/4G$, and the simple pole of $\zeta$ fixes the logarithmic coefficient
to the canonical value $\alpha=1$. Feeding this back through the correspondence
gives the closed-form geometry
\begin{equation}
f(r)=1-\frac{2GMr}{r^{2}+\lz^{2}},\qquad
\lz^{2}=\frac{\alpha G}{\pi},
\label{eq:intro_metric}
\end{equation}
whose properties we develop in detail: Reissner--Nordstr\"om horizon structure
without Coulombic hair, a positive-energy effective source of
vacuum-polarization type, a softened central singularity, a bounded temperature,
and a cold Planckian remnant. We stress that Eq.~\eqref{eq:intro_metric} is not
put in by hand. The same length $\lz$ comes back independently from demanding
that $dM=TdS$ hold exactly with the logarithmically corrected entropy. The
functional form belongs to the familiar family of regularized black
holes~\cite{Bardeen,Hayward,Nicolini,KazakovSolodukhin}, but here it is an
output rather than an ansatz, and its single parameter is fixed by the analytic
structure of $\zeta$ at $s=1$. The zeros re-enter only when the exact Euler
product is deformed, where they contribute log-periodic oscillations in the area
of relative size $e^{-3A/16G}$; within this framework the Riemann hypothesis
becomes the statement that these arithmetic corrections are as suppressed as
they can be.

The resulting geometry
is not astrophysically distinguishable from Schwarzschild: the relative correction at
the horizon is $\sim\!10^{-77}$ for a solar-mass black hole. The content of the
construction is structural and Planckian, and its value lies in being
falsifiable in three independent places, as discussed in Sec.~\ref{sec:disc}.

The paper is organized as follows.
Sections~\ref{sec:spectra}--\ref{sec:counting} establish that the exponential
degeneracy required by black-hole thermodynamics comes from the Euler product
and not from the zeros. Section~\ref{sec:map} fixes the geometric spectral map
and derives the corrected horizon entropy, including the coefficient of its
logarithmic term. Section~\ref{sec:solution} contains the main new result: an
explicit closed-form quantum-corrected black hole, with its horizon structure,
thermodynamics, effective source and evaporation endpoint.
Section~\ref{sec:zeros} isolates the role of the zeros, and
Secs.~\ref{sec:disc}--\ref{sec:conc} discuss the status of the assumptions and
conclude. We use $\hbar=c=k_B=1$ and $G=\ell_P^2=m_P^{-2}$.

\section{Two arithmetic spectra}
\label{sec:spectra}

We can read Eq.~\eqref{eq:zeta} as a partition function. Each prime $p$ is a bosonic
oscillator of energy $\epsilon_p=\ln p$, the occupation number $a_p$ counts how
many quanta sit in that mode, and a multiparticle state is labelled by the
integer $n=\prod_p p^{a_p}$ with total energy $E_n=\ln n$. Unique factorization
guarantees that this labelling is one to one: every integer appears exactly
once, with no degeneracy. With $\beta$ the inverse temperature, the Euler
product is then literally the partition function of this free gas,
\begin{equation}
Z_\ar(\beta)=\prod_p\left(1-e^{-\beta\ln p}\right)^{-1}=\zeta(\beta),
\label{eq:euler}
\end{equation}
and the usual canonical relations $U=-\partial_\beta\ln Z$ and
$S=\ln Z-\beta\partial_\beta\ln Z$ give
\begin{equation}
U=-\frac{\zeta'(\beta)}{\zeta(\beta)},\qquad
S_\ar=\ln\zeta(\beta)+\beta U .
\label{eq:canonical}
\end{equation}

The counting is simplest in the microcanonical ensemble. The states with energy
below $E$ are the integers below $e^{E}$, so
$\Omega_\ar(E)=\lfloor e^{E}\rfloor$ and the smoothed density of states is
$\rho_\ar(E)=d\Omega_\ar/dE=e^{E}$, up to fluctuations of order unity. The
entropy is therefore linear in the energy,
\begin{equation}
S_\ar(E)=\ln\Omega_\ar(E)\simeq \frac{E}{T_H},\qquad T_H=1 ,
\label{eq:hagedorn}
\end{equation}
in units in which $\epsilon_p=\ln p$. Linearity is the whole point. A spectrum
whose entropy is proportional to its energy has a single, energy-independent
temperature, and here that temperature is $T_H=1$. The same result follows in
the canonical ensemble, together with its subleading term, in
Sec.~\ref{sec:logcoeff}.

Physically, $T_H$ is a limiting temperature. The number of states grows exactly
as fast as the Boltzmann factor $e^{-\beta E}$ decays, so
$Z_\ar=\sum_n e^{-\beta\ln n}$ diverges as $\beta\to1^{+}$: heating the gas
towards $\beta=1$ pumps energy into creating new states rather than into raising
the temperature. This divergence is nothing other than the simple pole of
$\zeta$ at $\beta=1$, and it is the standard signature of Hagedorn
behavior~\cite{Hagedorn,AtickWitten}. The same statement can be made at the
single-particle level using the prime number theorem, which gives the mode
density
\begin{equation}
\rho_{\rm prime}(\epsilon)=\frac{d}{d\epsilon}\pi(e^{\epsilon})
\simeq\frac{e^{\epsilon}}{\epsilon} ,
\label{eq:primedensity}
\end{equation}
exponential up to a logarithm. Note that $E$ is a pure number here, so $T_H$ is
dimensionless; a physical scale enters only through the map of
Sec.~\ref{sec:map}.

The second spectrum is built from the nontrivial zeros. Assuming
Hilbert--P\'olya, we take the ordinates $t_n$ to be energy levels, $E_n\equiv
t_n$, and define $Z_\zeta(\beta)=\sum_n e^{-\beta t_n}$. The Riemann--von
Mangoldt counting formula,
$N_\zeta(T)=(T/2\pi)\ln(T/2\pi)-T/2\pi+7/8+O(\ln T)$, gives the smooth level
density
\begin{equation}
\bar\rho_\zeta(E)\simeq\frac{1}{2\pi}\ln\!\left(\frac{E}{2\pi}\right).
\label{eq:zerodensity}
\end{equation}
This grows only logarithmically. The contrast with
Eq.~\eqref{eq:primedensity} could hardly be sharper: the zeros are an
exponentially sparser spectrum than the primes. As a consequence $Z_\zeta$
converges for every $\beta>0$, and the zero gas has no limiting temperature at
all.

The two spectra are not independent, but they enter in different ways. The
oscillatory part of the level density of the zeros is controlled by the primes
through $-\zeta'/\zeta=\sum_n\Lambda(n)n^{-s}$, exactly as in a semiclassical
trace formula whose primitive periodic orbits have lengths $L_{p^m}=m\ln p$.
The dictionary is zeros $\leftrightarrow$ energy levels, primes
$\leftrightarrow$ periodic orbits. Which of the two carries the thermodynamics
is the question we turn to next.

\section{What black-hole thermodynamics requires}
\label{sec:counting}

A horizon of area $A$ is to be described by a Hilbert space of dimension
$e^{A/4G}$. Suppose that the horizon makes available the arithmetic states with
energies below some cutoff $E_c(A)$. The Bekenstein--Hawking area law is assumed, provided that
\begin{equation}
S_\ar\big(E_c(A)\big)=\ln\Omega_\ar\big(E_c(A)\big) =\frac{A}{4G}.
\label{eq:criterion}
\end{equation}
This is the only place where gravity enters the counting; everything else is
arithmetic. Two demands follow. First, the area is the extensive variable of the
horizon, so a cutoff growing linearly in $A$ has to produce an entropy linear in
$A$: the spectrum must have an \emph{extensive} entropy, $S\propto E$. Second,
the cutoff must be geometrically natural rather than fitted---a requirement made
precise in Sec.~\ref{sec:map}, where the Schwarzschild limit fixes $E_c(A)$
uniquely.

The prime gas passes the first test immediately. Equation~\eqref{eq:hagedorn}
gives $S_\ar=E/T_H$, so Eq.~\eqref{eq:criterion} returns
\begin{equation}
E_c\simeq T_H\,\frac{A}{4G},
\label{eq:Ecprelim}
\end{equation}
linear in the area, as required. The reason is the exponential mode density
\eqref{eq:primedensity}: only a Hagedorn spectrum converts a cutoff linear in the
area into an entropy linear in the area. Equivalently, the integers themselves
are counted exponentially, $\Omega_\ar(E)=\lfloor e^{E}\rfloor$, which is
precisely the structure that $e^{A/4G}$ demands.

One can now argue and see how the zeros fail to reproduce such a result. At the one-particle level the failure is
immediate: Eq.~\eqref{eq:zerodensity} gives
$N_\zeta(E)\simeq(E/2\pi)\ln(E/2\pi)$, so a single-particle count gives
$\ln N_\zeta\simeq\ln E+\ln\ln E$, and matching Eq.~\eqref{eq:criterion} would
require a cutoff $E_c\sim e^{A/4G}$, which is neither geometric nor bounded in
any useful sense. The natural repair is to build a Fock space on the zeros,
which is what the statement ``the zeros are the levels of a quantum system''
means thermodynamically. Writing $\ln Z_\zeta=-\sum_n\ln(1-e^{-\beta t_n})$ and
using the smooth density \eqref{eq:zerodensity},
\begin{equation}
\ln Z_\zeta(\beta)\simeq\int^{\infty} d\epsilon\;\bar\rho_\zeta(\epsilon)
\left[-\ln\!\left(1-e^{-\beta\epsilon}\right)\right] .
\end{equation}
Rescaling $u=\beta\epsilon$ and using
$\int_0^\infty du\,[-\ln(1-e^{-u})]=\zeta(2)=\pi^2/6$, the small-$\beta$
behavior is
\begin{equation}
\ln Z_\zeta(\beta)\simeq\frac{\pi\,\mathcal L}{12\,\beta},
\qquad \mathcal L\equiv\ln\frac{1}{2\pi\beta} ,
\end{equation}
whence $E\simeq\pi\mathcal L/12\beta^{2}$ and
\begin{equation}
S_\zeta(E)\simeq\sqrt{\frac{\pi\,\mathcal L\,E}{3}}
\simeq\sqrt{\frac{\pi}{6}\,E\ln E} .
\label{eq:zeroentropy}
\end{equation}
This is a Hardy--Ramanujan law of Cardy type. The entropy of the
second-quantized zero gas grows as $\sqrt E$ up to logarithms, because the
one-particle density \eqref{eq:zerodensity} grows logarithmically rather than
exponentially. Accordingly $Z_\zeta$ is finite for every $\beta>0$, there is no
Hagedorn pole, and $S_\zeta/E\to0$: the thermodynamics is not extensive.

The consequence for Eq.~\eqref{eq:criterion} is quantitative. With the
geometrically selected cutoff $E_c\propto A$ of Sec.~\ref{sec:map}, the zero gas
yields $S\propto\sqrt{A\ln A}$ rather than $A$. Forcing $S=A/4G$ instead
requires $E_c\propto A^{2}/\ln A\propto M^{4}$, which is not what the
entropy--geometry map returns and which does not reduce to Schwarzschild. The
mismatch is therefore not one of normalization but of growth rate, and it is
exponentially large.

It is worth checking how much of this depends on the statistics of the horizon
gas, and the answer is essentially none. In the bosonic gas each prime mode can
be occupied any number of times, $a_p\in\{0,1,2,\ldots\}$, giving
$Z_B(\beta)=\zeta(\beta)$. In the fermionic gas the exclusion principle allows
at most one particle per mode, $a_p\in\{0,1\}$, which selects the squarefree
integers and gives~\cite{Spector}
\begin{equation}
Z_F(\beta)=\prod_p\left(1+e^{-\beta\ln p}\right)
=\frac{\zeta(\beta)}{\zeta(2\beta)} .
\label{eq:ZF}
\end{equation}
Both partition functions have the same leading singularity, a simple pole at
$\beta=1$, and differ only in its residue: $1$ in the bosonic case and $6/\pi^2$
in the fermionic case. Since a simple pole gives $S=E+\ln E+\text{const}$, and
since the metric will turn out to depend only on $S'(r)$, the additive constant
never reaches the geometry. Bosonic and fermionic arithmetic gases therefore
produce the \emph{same} leading black-hole geometry \eqref{eq:intro_metric}. The
statistics affects only the exponentially small ripples of
Sec.~\ref{sec:zeros}, not the macroscopic spacetime. This universality is a
direct consequence of the fact that the pole of the zeta function, and not its
zeros, controls the extensive thermodynamics.

The two spectra thus play complementary roles, exactly as the explicit formula
suggests. The primes, in their role as periodic orbits, control the exponential
proliferation of states and hence the area law. The zeros, in their role as
levels, are far too sparse to carry it and can only modulate it---as they in
fact do, in Sec.~\ref{sec:zeros}.

\section{The spectral map and horizon entropy}
\label{sec:map}

\subsection{The map is fixed by the Schwarzschild limit}

Since Eq.~\eqref{eq:hagedorn} already supplies the required growth, the only
remaining freedom is the cutoff $E_c(A)$ specifying which arithmetic energies a
horizon of area $A$ makes available. Equations~\eqref{eq:hagedorn} and
\eqref{eq:criterion} give $E_c\simeq T_HA/4G$; we now show that this is not a
matching prescription but is selected geometrically.

For a static, spherically symmetric, asymptotically flat spacetime
\begin{equation}
ds^2=-f(r)dt^2+\frac{dr^2}{f(r)}+r^2d\Omega_2^2,
\end{equation}
the entropy--geometry correspondence~\cite{Anand,AnandTopo,JusufiAnand}
reconstructs the metric from a prescribed entropy of the sphere of radius $r$,
\begin{equation}
f(r)=1-\frac{4\pi M}{S'(r)} .
\label{eq:entropy_geometry}
\end{equation}
For $S=\pi r^2/G$ this returns Schwarzschild. Identifying
$S(r)=S_\ar(E_c(r))$ and using Eq.~\eqref{eq:hagedorn}, so that
$S'(r)=T_H^{-1}E_c'(r)$, the requirement that the infrared geometry be
Schwarzschild, $S'(r)=2\pi r/G$, forces $E_c'=2\pi T_Hr/G$ and hence
\begin{equation}
 E_c(A)=T_H\frac{A}{4G}
\label{eq:map}
\end{equation}
up to an additive constant, in agreement with Eq.~\eqref{eq:Ecprelim}. The
quadratic dependence $E_c\propto r^2\propto M^2$ is therefore geometrically
selected: identifying the arithmetic cutoff with the ADM mass instead would not
reproduce the Schwarzschild limit.

Two consistency conditions of the correspondence~\cite{Anand} will be used
below. Positivity of the reconstructed mass parameter requires $S'(r)>0$, and
since Eq.~\eqref{eq:entropy_geometry} gives $f'(r_+)=S''(r_+)/S'(r_+)$ at a
horizon, positivity of the temperature requires $S''(r_+)>0$. Both will hold
precisely on the outer branch of the solution obtained in
Sec.~\ref{sec:solution}.

\subsection{The logarithmic coefficient is universal}
\label{sec:logcoeff}

The subleading term in $S_\ar$ is fixed by the analytic structure of $\zeta$ at
its pole. Writing $x=\beta-1$ and using
$\zeta(\beta)=x^{-1}+\gamma+\mathcal O(x)$,
\begin{equation}
\ln Z_\ar=-\ln x+\gamma x,\qquad E=-\partial_\beta\ln Z_\ar=\frac1x-\gamma,
\end{equation}
so that $S=\ln Z+\beta E$ gives
\begin{equation}
 S_\ar(E)=\frac{E}{T_H}+\ln\!\frac{E}{T_H}+1+\mathcal O(E^{-1}).
\label{eq:logderived}
\end{equation}
The logarithm is not model dependent. Any partition function with a simple
Hagedorn pole, $Z\simeq \mathcal A/(\beta-\beta_H)$, yields
$S=\beta_HE+\ln E+\ln\mathcal A+1$, with the residue $\mathcal A$ entering only
through an additive constant. In particular the fermionic (squarefree) Riemann
gas \eqref{eq:ZF} has the same pole and therefore the same coefficient.
Combining Eqs.~\eqref{eq:map} and \eqref{eq:logderived},
\begin{equation}
S(A)=\frac{A}{4G}+\alpha\ln\!\frac{A}{4G}+\text{const},
\qquad \alpha=1 .
\label{eq:Scorrected}
\end{equation}
Because Eq.~\eqref{eq:entropy_geometry} involves only $S'(r)$, the additive
constant---the most scheme-dependent piece---drops out of the geometry, and
$\alpha$ is the only quantity transmitted to the metric.

We note that $\alpha=1$ is the canonical value: the logarithm originates in
energy fluctuations, and a strictly microcanonical count, for which
$\Omega_\ar(E)=\lfloor e^E\rfloor$ exactly, gives $\alpha=0$. This mirrors the
known ensemble sensitivity of logarithmic corrections to
$S_\BH$~\cite{KaulMajumdar,Carlip,Das,Sen,Solodukhin}, and we therefore keep
$\alpha$ explicit below so that its effect on the geometry can be read off
directly. The \emph{sign} of $\alpha$, however, is not a matter of convention
here. As shown in Sec.~\ref{sec:solution}, $\alpha>0$ produces a regular
two-horizon geometry with a positive-energy source, whereas $\alpha<0$ would
give $\lz^2<0$ and a naked singular surface at $r=|\lz|$ where $f$ diverges. The
canonical prime gas sits on the physically admissible side.

\section{The arithmetic black hole}
\label{sec:solution}

\subsection{The solution}

\begin{figure}[t]
\includegraphics[width=\columnwidth]{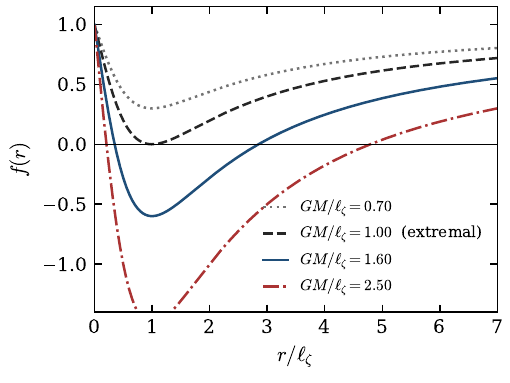}
\caption{The metric function \eqref{eq:metric} in units $\lz=G=1$.
Below $GM=\lz$ the geometry is horizonless; at $GM=\lz$ the two
horizons of Eq.~\eqref{eq:horizons} merge into the degenerate horizon
$r_*=\lz$ of the extremal remnant; above it the structure is that of
Reissner--Nordstr\"om with $q^2\to\lz^2$.}
\label{fig:metric}
\end{figure}

Inserting Eq.~\eqref{eq:Scorrected}, i.e.\
$S(r)=\pi r^2/G+\alpha\ln(\pi r^2/G)$, into
Eq.~\eqref{eq:entropy_geometry} with
$S'(r)=2(\pi r^2+\alpha G)/(Gr)$ gives a closed form, with no expansion
required:
\begin{equation}
f(r)=1-\frac{2GMr}{r^{2}+\lz^{2}},
\qquad
\lz^{2}=\frac{\alpha G}{\pi}.
\label{eq:metric}
\end{equation}
For $\alpha=1$ the arithmetic length is $\lz=\ell_P/\sqrt\pi\simeq
0.564\,\ell_P$; for $\alpha=0$ Schwarzschild is recovered exactly. The
statistics and ensemble of the horizon gas are thus imprinted directly on the
geometry. The metric function is displayed in Fig.~\ref{fig:metric}.

Asymptotically,
\begin{equation}
f(r)=1-\frac{2GM}{r}+\frac{2\alpha}{\pi}\frac{G^2M}{r^{3}}
+\mathcal O(r^{-5}),
\label{eq:asymptotic}
\end{equation}
so there is no $1/r^2$ term: the arithmetic correction carries no Coulombic hair
and is invisible to any asymptotic charge measurement. The leading correction
has the same structure as the one-loop quantum correction to the Schwarzschild
metric obtained from graviton loops~\cite{Donoghue,BjerrumBohr}, with the
coefficient here fixed by $\alpha$.

The solution is not merely consistent with the corrected thermodynamics at
leading order: it satisfies the first law identically. To see this, consider the
two-parameter family $f=1-2GMr/(r^{2}+\ell^{2})$ with $\ell$ left free. The
horizon follows from $f(r_+)=0$, i.e.\ $r_+^2 - 2GMr_+ + \ell^2=0$, so
$M=(r_+^2+\ell^2)/(2Gr_+)$. The Hawking temperature is
\begin{equation}
T = \frac{r_+^2 - \ell^2}{4\pi r_+(r_+^2+\ell^2)},
\end{equation}
while the entropy from the correspondence is
$S(r)=\pi r^2/G+\alpha\ln(\pi r^2/G)$. Computing the first-law derivative,
\begin{equation}
\frac{dM}{dr_+} - T\frac{dS}{dr_+}
= \frac{(r_+^2-\ell^2)(\pi\ell^2-\alpha G)}
        {2\pi G r_+^2(r_+^2+\ell^2)},
\end{equation}
which vanishes for all $r_+$ if and only if
\begin{equation}
\ell^{2}=\frac{\alpha G}{\pi}=\lz^{2}.
\label{eq:firstlawfix}
\end{equation}
The arithmetic length is therefore fixed twice over: once by the
entropy--geometry reconstruction, and independently by exact consistency of
$dM=TdS$ with the logarithmically corrected entropy. Equivalently, the
perturbative ansatz $f=1-2GM/r+\lambda G^2M/r^3$ subjected to the first law
returns $\lambda=2\alpha/\pi$, matching Eq.~\eqref{eq:asymptotic}.

\subsection{Horizons and the remnant}

Equation~\eqref{eq:metric} vanishes where $r^2-2GMr+\lz^2=0$, so the horizon
polynomial coincides with that of Reissner--Nordstr\"om under
$q^2\to\lz^2$:
\begin{equation}
r_\pm=GM\pm\sqrt{G^2M^2-\lz^{2}} .
\label{eq:horizons}
\end{equation}
A macroscopic horizon is only mildly displaced, $r_+\simeq2GM-\alpha/2\pi M$,
but the structure at the Planck scale is qualitatively new. Horizons exist only
for
\begin{equation}
 M\ge M_*=\frac{\lz}{G}=\sqrt{\frac{\alpha}{\pi}}\;m_P
\simeq0.56\,m_P ,
\label{eq:remnant}
\end{equation}
with a degenerate horizon at $r_*=\lz$ when $M=M_*$. The extremal configuration
has $A_*=4\alpha G$ and hence $A_*/4G=\alpha$: for $\alpha=1$ the remnant
carries a single nat of horizon entropy. Below $M_*$ the geometry is
horizonless. We stress that Eq.~\eqref{eq:Scorrected} was derived as a large-$A$
expansion, so Eq.~\eqref{eq:remnant} indicates the scale at which the horizon
terminates rather than a precise mass; the sign of the effect, however, follows
from $\alpha>0$ alone.

\subsection{Thermodynamics}

From $\kappa=f'(r_+)/2$, $T=\kappa/2\pi$, and $r_+^2+\lz^2=2GMr_+$,
\begin{equation}
T=\frac{r_+^{2}-\lz^{2}}{4\pi r_+\left(r_+^{2}+\lz^{2}\right)},
\qquad
M=\frac{r_+^{2}+\lz^{2}}{2Gr_+} ,
\label{eq:temperature}
\end{equation}
which recovers $T\to1/8\pi GM$ for $r_+\gg\lz$. This is exactly
$T=S''(r_+)/4\pi S'(r_+)$, and since $S''(r)=2(\pi r^2-\alpha G)/(Gr^2)$ is
positive precisely for $r>\lz$, the positive-temperature condition of
Sec.~\ref{sec:map} selects the outer branch $r_+>\lz$ of
Eq.~\eqref{eq:horizons}. The consistency conditions of the correspondence and
the horizon structure of the solution therefore agree without further input.

\begin{figure}[t]
\includegraphics[width=\columnwidth]{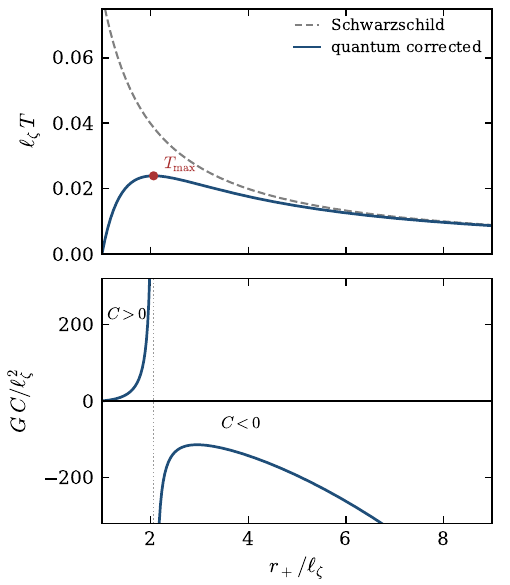}
\caption{Upper panel: the Hawking temperature \eqref{eq:temperature} (solid)
against the Schwarzschild value $1/4\pi r_+$ (dashed). The temperature is
bounded by $T_{\max}\simeq0.0239/\lz$ at
$r_+=\sqrt{2+\sqrt5}\,\lz$ and vanishes at extremality. Lower panel: the
heat capacity \eqref{eq:heatcap}, negative in the Schwarzschild-like regime and
positive on the near-extremal branch, with the divergence at the temperature
maximum marking the transition (dotted line).}
\label{fig:thermo}
\end{figure}

The temperature is bounded. It reaches a maximum
$T_{\max}\simeq0.0239/\lz\simeq0.042\,m_P$ at
$r_+=\sqrt{2+\sqrt5}\,\lz\simeq2.06\,\lz$, and vanishes at extremality. With
\begin{equation}
\frac{dM}{dr_+}=\frac{r_+^{2}-\lz^{2}}{2Gr_+^{2}},
\qquad
\frac{dT}{dr_+}=-\,\frac{r_+^{4}-4r_+^{2}\lz^{2}-\lz^{4}}
{4\pi r_+^{2}\left(r_+^{2}+\lz^{2}\right)^{2}},
\label{eq:dMdr}
\end{equation}
the heat capacity is found to be
\begin{equation}
C=-\,\frac{2\pi\left(r_+^{2}-\lz^{2}\right)
\left(r_+^{2}+\lz^{2}\right)^{2}}
{G\left(r_+^{4}-4r_+^{2}\lz^{2}-\lz^{4}\right)} .
\label{eq:heatcap}
\end{equation}
It is negative for $r_+>\sqrt{2+\sqrt5}\,\lz$, where it reproduces the
Schwarzschild result $C\simeq-2\pi r_+^2/G=-8\pi GM^2$, and positive below;
both branches are shown in Fig.~\ref{fig:thermo}. Hawking evaporation therefore
proceeds normally, slows down as $T$ turns over, and ends on a cold extremal
remnant of mass $M_*$ rather than in a divergent burst---the standard remnant
scenario~\cite{Nicolini}, obtained here from arithmetic input rather than
postulated. The presence of an inner horizon $r_-$ carries the usual caveat of
Cauchy-horizon instability, which the present static analysis does not address.

\subsection{Effective source and central structure}

Writing $f=1-2Gm(r)/r$ gives the quasilocal mass function
$m(r)=Mr^{2}/(r^{2}+\lz^{2})$, which interpolates between $m(0)=0$ and
$m(\infty)=M$. The Einstein equations then define an effective anisotropic fluid
with
\begin{equation}
\rho=\frac{m'}{4\pi r^{2}}
=\frac{M\lz^{2}}{2\pi r\left(r^{2}+\lz^{2}\right)^{2}},
\qquad p_r=-\rho ,
\end{equation}
and tangential pressure $p_t=-\rho-r\rho'/2$. The energy density is positive
everywhere, falls off as $M\lz^2/2\pi r^{5}$ at large $r$, and integrates to the
full ADM mass. The null energy condition holds, since $\rho+p_r=0$ and
$\rho+p_t=-r\rho'/2>0$ for monotonically decreasing $\rho$. The source is thus
of vacuum-polarization type, with support concentrated within a few $\lz$ of the
center.

The center remains singular but is substantially softened. The Ricci scalar,
\begin{equation}
R=\frac{4GM\lz^{2}\left(3\lz^{2}-r^{2}\right)}
{r\left(r^{2}+\lz^{2}\right)^{3}}
\;\xrightarrow[r\to0]{}\;\frac{12GM}{\lz^{2}\,r},
\end{equation}
diverges only as $r^{-1}$, and the Kretschmann scalar behaves as
\begin{equation}
K\;\xrightarrow[r\to0]{}\;\frac{32G^{2}M^{2}}{\lz^{4}\,r^{2}},
\end{equation}
to be compared with the Schwarzschild value $K=48G^2M^2/r^{6}$. The arithmetic
correction does not remove the singularity---one should not expect a
leading-order entropy correction to do so---but it reduces its degree by four
powers.

The relative correction to the metric at the horizon is
$\lz^2/r_+^2\simeq\alpha(m_P/2M)^2/\pi$, i.e.\ $\sim10^{-77}$ for a solar-mass
black hole and $\sim10^{-41}$ for a $10^{15}\,$g primordial one. The solution is
therefore observationally indistinguishable from Schwarzschild in any
astrophysical setting; its content is the Planck-scale endpoint,
Eqs.~\eqref{eq:remnant} and \eqref{eq:temperature}, not a shadow or ringdown
signature. This is a feature rather than a defect: the framework makes no claim
that arithmetic structure is astrophysically visible.

\section{Where the zeros enter and how they are avoided in the metric}
\label{sec:zeros}

The bosonic prime gas counts the integers exactly, $\rho_\ar(E)=e^{E}$, and the
zeros drop out of the result: the leading area law is insensitive to the Riemann
hypothesis. The cancellation is easy to see. The density of states follows from
the inverse Laplace transform $\rho(E)=\mathcal L^{-1}[\zeta(\beta)]$, and the
nontrivial zeros are zeros of $\zeta(\beta)$, not poles, so they contribute no
residue to the contour integral. The only contribution comes from the pole at
$\beta=1$, which gives $\rho(E)=e^{E}$ up to exponentially small corrections. In
the bosonic gas the zeros are therefore absent from the leading and even the
subleading thermodynamics.

They reappear as soon as the exact Euler product is deformed. For the fermionic
(squarefree) gas, $Z_F(\beta)=\zeta(\beta)/\zeta(2\beta)$, the zeros
$\rho_n=\sigma_n+it_n$ of $\zeta(2\beta)$ become poles of $Z_F$ at
\begin{equation}
\beta_n=\frac{\rho_n}{2}=\frac{\sigma_n}{2}+\frac{it_n}{2},
\label{eq:polelocation}
\end{equation}
the halving being the only place where the argument doubling of $\zeta(2\beta)$
enters. Inverse Laplace transformation then gives
\begin{equation}
\rho_F(E)=\frac{6}{\pi^{2}}e^{E}
+\sum_{t_n>0} a_n\,e^{\sigma_n E/2}
\cos\!\left(\frac{t_nE}{2}+\varphi_n\right)+\cdots,
\label{eq:rhoF}
\end{equation}
where the sum runs over zeros in the upper half plane and
$a_ne^{i\varphi_n}=\zeta(\rho_n/2)/\zeta'(\rho_n)$, the factor of two from the
residue at $\beta=\rho_n/2$ being cancelled by the pairing with the complex
conjugate zero. We deliberately keep $\sigma_n$ unspecified: the Riemann
hypothesis is not used anywhere in what follows, and enters only at the very
end, as the statement that fixes its value.

Since the oscillatory piece of Eq.~\eqref{eq:rhoF} is exponentially small
relative to the leading term, the corresponding entropy correction is
$\delta S=\ln\!\big(1+\rho_{\rm osc}/\rho_{\rm lead}\big)\simeq
\rho_{\rm osc}/\rho_{\rm lead}$, controlled by the ratio of the two residue
exponentials,
\begin{equation}
\frac{e^{\sigma_n E/2}}{e^{E}}=e^{-\left(1-\sigma_n/2\right)E}.
\label{eq:ratio}
\end{equation}
Under the map \eqref{eq:map}, $E=A/4G$, the horizon entropy therefore acquires
log-periodic oscillations in the area,
\begin{equation}
\delta S(A)\sim \sum_{t_n>0} a_n\,
e^{-\left(1-\frac{\sigma_n}{2}\right)\frac{A}{4G}}
\cos\!\left(\frac{t_nA}{8G}+\varphi_n\right),
\label{eq:oscillations}
\end{equation}
whose frequencies are set by the imaginary parts of the zeros and whose
envelopes are set by their real parts.

These entropy oscillations propagate into the metric through the
entropy--geometry correspondence. Let $S(r)=S_0(r)+\delta S(r)$, where
$S_0(r)=\pi r^2/G$ is the uncorrected entropy. Substituting into
Eq.~\eqref{eq:entropy_geometry} and expanding to first order in $\delta S$
gives
\begin{equation}
f(r)=1-\frac{4\pi M}{S_0'(r)}
+\frac{4\pi M}{(S_0'(r))^2}\delta S'(r) + \mathcal O(\delta S^2).
\end{equation}
Using $S_0'(r)=2\pi r/G$, the correction to the metric function is
\begin{equation}
\delta f(r)=\frac{G^{2}M}{\pi r^{2}}\,\delta S'(r).
\label{eq:deltaf}
\end{equation}
This is the exact relation between the zero-induced entropy oscillations and the
metric perturbation. Inserting Eq.~\eqref{eq:oscillations} with $A=4\pi r^2$,
so that the envelope becomes $\exp[-(1-\sigma_n/2)\pi r^2/G]$ and
$\Xi_n=t_n\pi r^2/2G+\varphi_n$, yields
\begin{eqnarray}\notag
\delta f(r)\sim  -\frac{GM}{r}
\sum_{t_n>0} &a_n&
e^{-\left(1-\frac{\sigma_n}{2}\right)\frac{\pi r^{2}}{G}}
\Big[\left(2-\sigma_n\right)\cos\Xi_n\\
&+&t_n\sin\Xi_n \Big].
\label{eq:deltaf_osc}
\end{eqnarray}
The real part of each zero thus controls both the envelope and the relative
weight of the two quadratures; the familiar coefficient $\tfrac32$ is the value
of $2-\sigma_n$ on the critical line.

It remains to ask how small these corrections can be. The zeros enter at
relative order $e^{-(1-\sigma_n/2)A/4G}$, and the least suppressed ripple is the
one with the largest real part, so the whole sum is governed by
\begin{equation}
\Theta\equiv\sup_n\sigma_n,\qquad
\delta f/f\big|_{\rm max}\sim
\exp\!\left[-\left(1-\frac{\Theta}{2}\right)\frac{A}{4G}\right].
\label{eq:theta}
\end{equation}
Larger $\Theta$ means \emph{weaker} suppression. Two facts bound $\Theta$ from
either side. The Euler product and the prime number theorem confine the
nontrivial zeros to the critical strip $0<\sigma_n<1$, so the exponent
$1-\Theta/2$ lies in $(\tfrac12,1)$ and the arithmetic ripples are always
exponentially small in the area---the geometry \eqref{eq:metric} is never
destabilised, whatever the zeros do. The functional equation
$\xi(s)=\xi(1-s)$, on the other hand, pairs every zero $\rho_n$ with $1-\rho_n$,
so the set $\{\sigma_n\}$ is symmetric about $\tfrac12$ and therefore
\begin{equation}
\Theta\ge\tfrac12,\qquad\text{hence}\qquad
1-\frac{\Theta}{2}\le\frac{3}{4}.
\label{eq:bound}
\end{equation}
The exponent is thus bounded above, and the bound is saturated precisely when
$\Theta=\tfrac12$, that is, when all nontrivial zeros lie on the critical line.
In that case Eqs.~\eqref{eq:oscillations} and \eqref{eq:deltaf_osc} collapse to
the single common envelope
\begin{equation}
\delta S(A)\sim e^{-3A/16G}\sum_{t_n>0}a_n
\cos\!\left(\frac{t_nA}{8G}+\varphi_n\right),
\label{eq:RHcase}
\end{equation}
and 
\begin{equation}
\delta f(r)\sim-\frac{GM}{r}e^{-3\pi r^{2}/4G}\sum_{t_n>0}a_n
\left[\tfrac32\cos\Xi_n+t_n\sin\Xi_n\right],
\end{equation}
and $e^{-3A/16G}$ is the fastest decay the arithmetic corrections can possibly
have. A single exceptional zero at, say, $\sigma=0.6$ would already loosen the
envelope to $e^{-0.7A/4G}$, and the corresponding ripples would be
exponentially larger at every horizon area. Within this framework, then, the
Riemann hypothesis is not an input but an extremal statement: it is exactly the
condition that arithmetic corrections to the black-hole area law are maximally
suppressed. We stress that this reading applies to the deformed (squarefree)
counting, where the zeros acquire residues at all; in the undeformed bosonic
gas they contribute nothing, and the leading geometry is independent of the
hypothesis altogether.

The zeros are thus \emph{avoided} in the leading geometry \eqref{eq:metric}
because they enter only as exponentially small perturbations of the entropy and
of the metric, with an envelope that the functional equation forces to be
narrowest on the critical line. The prime sector, through the simple pole at
$\beta=1$, completely dominates the extensive thermodynamics and fixes the
closed-form metric. The zeros merely decorate that geometry with unobservably
small quantum ripples.

\section{Discussion}
\label{sec:disc}

It is worth separating what is derived from what is assumed. Derived: the
Hagedorn structure of $Z_\ar=\zeta(\beta)$ and Eq.~\eqref{eq:logderived}; the
sub-extensive entropy \eqref{eq:zeroentropy} of the second-quantized zero gas,
which excludes it as a carrier of the area law; the map \eqref{eq:map} given the
Schwarzschild limit; and, given Eq.~\eqref{eq:entropy_geometry}, the solution
\eqref{eq:metric} together with all of Sec.~\ref{sec:solution}. Assumed: that
the horizon degrees of freedom are the arithmetic ones at all,  the validity of
Eq.~\eqref{eq:entropy_geometry} as a reconstruction principle; and the
Hilbert--P\'olya interpretation, used only in Secs.~\ref{sec:counting}
and~\ref{sec:zeros}.

The framework is falsifiable in three independent places. First, $\alpha$ is
predicted, not fitted: a microscopic derivation of horizon entropy giving
$\alpha\neq1$ (for instance the frequently quoted
$-3/2$~\cite{KaulMajumdar,Das}) is incompatible with the canonical prime gas,
and would force either a different ensemble or a different arithmetic sector. A
negative $\alpha$ would in addition be incompatible with the existence of the
regular two-horizon geometry itself. Second, the correction is required to have
no $1/r^2$ term, Eq.~\eqref{eq:asymptotic}; an arithmetic origin cannot produce
Coulombic hair. Third, the residual structure is required to be log-periodic in
$A/4G$ with frequencies $t_n/2$, Eq.~\eqref{eq:oscillations}, rather than a
power series.

The most significant open problem remains the first link of the chain. Nothing
above explains why a horizon should make available precisely the arithmetic
energies below $E_c=T_HA/4G$; that relation was extracted from the Schwarzschild
limit rather than derived from a horizon operator. Related to this, the fact
that the horizon gas sits exactly at its Hagedorn point---which is what makes
the entropy extensive in $E$ and hence linear in $A$---is reminiscent of the
string/black-hole correspondence principle~\cite{HorowitzPolchinski}, where the
transition occurs when the string temperature reaches $T_H$. Whether that
resemblance can be made structural rather than suggestive is the natural next
question. Until such an operator is constructed, the solution \eqref{eq:metric}
should be read as the geometry that arithmetic state counting implies, not as
evidence that arithmetic state counting is correct.

\section{Conclusions}
\label{sec:conc}

We have shown that the arithmetic statistical mechanics of the Euler product,
combined with the entropy--geometry correspondence, produces a definite
black-hole solution rather than an analogy. The exponential degeneracy demanded
by $S_\BH$ is supplied by the primes, whose mode density grows as
$e^{\epsilon}/\epsilon$, and not by the Riemann zeros, which are too sparse by
an exponential margin even after second quantization, their Fock-space entropy
growing only as $\sqrt{E\ln E}$. The simple pole of $\zeta$ at $s=1$ then fixes
the subleading entropy universally, $\alpha=1$, and the resulting geometry is
$f(r)=1-2GMr/(r^{2}+\lz^{2})$ with $\lz^{2}=\alpha G/\pi$, a value returned
independently by exact consistency of the first law.

This spacetime has Reissner--Nordstr\"om horizon structure without Coulombic
hair, a positive-energy anisotropic effective source obeying the null energy
condition, a central singularity softened from $K\sim r^{-6}$ to $K\sim r^{-2}$,
a bounded Hawking temperature, and an extremal cold remnant of mass
$M_*\simeq0.56\,m_P$ that terminates the evaporation. The nontrivial zeros
survive only as log-periodic corrections of relative size $e^{-3A/16G}$, which
makes the Riemann hypothesis, in this setting, a statement about the maximal
suppression of arithmetic corrections to the area law.

The decisive open step is unchanged: to construct the horizon operator whose
arithmetic cutoff scales with the area. What has changed is the target. A
candidate construction must now reproduce not only $A/4G$ but a specific metric,
a specific remnant mass, and a specific pattern of residual oscillations.

\begin{acknowledgments}
Ankit Anand is financially supported by the Institute's postdoctoral fellowship at IIT Kanpur.
\end{acknowledgments}

\bibliographystyle{ref}
\bibliography{ref}
\end{document}